# Chapter 3: The History and Future Prospects of Open Data and Open Source Software

Feras A. Batarseh, Abhinav Kumar, Sam Eisenberg

To the past, or to the future, to an age where thought is free, from the age of big brother, from the age of the thought police, from a dead man: Greetings.  *1984, by George Orwell*

**Abstract:** Open data for all New Yorkers – that is the tagline on New York City's open data website. Open government is being promoted at most countries of the western world. Governments' transparency levels are being measured by the amount of data they share through their online public repositories. Additionally, open source software is promoted at governments, academia, and the industry – this is the new digital story of this century, and the new testament between the Gods of technology and there users; data and software openness will redefine the path forward, and aim to rekindle our collective intelligence. Data and Software Openness can re-define Data Democracy, and be the catalyst for its progress. This chapter provides a historical insight into data and software openness, the beginnings, the heroes, prospects for the future, and all things we cannot afford to negotiate or lose.

## 1.1 Introduction to the History of Open-Source

Open-Source Software (OSS) has had a lasting impact on business, government institutions, and scientific research. OSS was initially adopted by and for developers and programming hobbyists.

OSS have then expanded to operating systems (most famously, Linux), virtual machines (such as Oracle's VM VirtualBox), and internet clients such as Mozilla Firefox [1]. The history of Open-Source (OS) begins from the 1950s [2]. Heavily relying on government funding, computing research was relegated to academia. According to David Berry, Professor of Digital Humanities at the University of Sussex, "*difficulties faced by computer scientists...and their early links to academic and scholarly norms of research...contributed to sharing practices that later fed directly into the ethics of the [Open-Source movement]*". From these origins, 'hackers', ranging from academics to electronics hobbyists, banded in various groups (e.g., MIT's Artificial Intelligence Club and the Homebrew Computing Club) aiming to "*overcome limitations of software systems [and] achieve novel outcomes*" [3].

In the 70s, the hacker culture began to decline as control over "*freely sharing files, tools and information*" began to be asserted by corporations [4]. Activists like Richard Stallman would crusade against the privatization of software, denouncing proprietary software as "*evil, because it required people to agree not to share and that made society ugly*" [5]. Referring to his peers who decided to join for-profit firms as traitors!

Stallman would create an incomplete GNU Operating System. Later, Linus Torvalds would develop and publicly release a Linux kernel for GNU, making it the first publicly available operating system. The release of GNU/Linux unleashed a wave of volunteer peer-to-peer collaborations that define open source development to this day. Differing from Stallman's ideological viewpoint, Torvalds reflected a more practical approach, stating, "I don't like single-issue people, why should business, which fuels so much of society's technological advancement, be excluded?" [5]. Eventually, Torvalds emphasis on Open-Source software was: "*the pragmatic goal of getting people to collaborate in order to create software more effectively*". That philosophy became dominant over Stallman's free software ideology [5]. His (Stallman's) evangelical crusade did, however, provide the legal framework for OS to exist upon. Stallman's *Free Software Foundation* formalized user rights and the copyleft licensing model, legally ensuring liberties that fundamentally differ from proprietary software, namely:

1. The right to full access to the source code
2. The right for anyone to run the program without restriction
3. The right to modify the source code
4. The right to distribute both the original software and the modified software
5. The right for users to know about their open source rights
6. The obligation to distribute derivatives under copyleft

These defined rights [1] were essential to ensuring the growth and stability of the OS community as corporations patented and trademarked software [6]. OSS's early relationship with industry would come to define many developments.

**1.2 Open-Source Software's Relationship to Corporations**

Historically, large corporations have had a tumultuous relationship with OSS [7]. Viewed as a source of competition, Microsoft CEO Steve Ballmer referred to Linux as a "malignant cancer" [8]. As a result, much of OSS was first developed by universities and computer hobbyists. OS usage has now become ubiquitous within private industry. A survey conducted by Black Duck Software showed the "*percentage of companies running part or all of their operations on Open-Source software had almost doubled between 2010 and 2015, from 42% to 78%*" [7]. Since then, corporations and governments not only incorporate OSS, but contribute to OS projects. The number of firms that contribute to open source projects rose from 50% in 2014 to 64% in 2016, signifying that the private sector will continue to define OS development heavily in the future [7].

Economic and technological necessity has led corporations to embrace Open-Source use and/or development. For example, by *open-sourcing* their software, companies "*establish their technology as de facto standards*"; profiting by offering related products and services [9]. Security is also another concern; proprietary/closed source software often lacks stability. According to a Google Security Survey, Microsoft IIS Web servers were found to be "*twice as likely to distribute malware*" as Open Source Apache Web servers [10]. Considering the number of variables constituting the likelihood of an attack, it is difficult to statistically assert such claims.

Experts like Rob Nash, senior lecturer in Computing at the University of Washington, are skeptical of the metrics used in these kinds of studies, stating: "*there are many variables besides the number of attacks against or reported vulnerabilities in comparable open source and proprietary software*". Regardless, OSS gives opportunities to more developers–novice and experts–to possibly fix security flaws and bugs in the code. This is vital in an age where "95% of software bugs are caused by 19 common and well-understood programming errors" [11]. As a result, there is an established opinion that OSS "*does not pose any significant barriers to security, but rather reinforces sound security practices by involving many people that expose bugs quickly*" [11]. A changed industry outlook towards OSS has coincided with the popularization of Artificial Intelligence (AI) and Data Science tools. The intersection of OS and data science is discussed in the next section of the chapter.

## 1.3 Open-Source Data Science Tools

The cost of data storage has declined drastically within the last ten years, a 15-20% cost reduction just in the past several years. Additionally, with cloud services, it is becoming feasible for companies to cheaply store massive data sets. Interpreting interrelated variables on a large scale is impossible to do without data science/analytic tools [12]. Data Science is defined as "*the study of the generalizable extraction of knowledge from data*" [13]. Specializing in areas such as computer science, mathematics, or statistics, data scientists extract useful information from a data set. Interest in data science and its applications have grown immensely, according to Google Trends. OS and proprietary tools for data science differ in several key ways. First, proprietary software is usually not as customizable, and if a defect is detected, it cannot easily be changed [14]. Second, because of OSS's ability to change quickly, OSS tends to be reliable and technically advanced [15]. Third, OSS's low/free cost, technical sophistication, and flexibility makes OSS an attractive option for researchers, whom usually add more features to the software [16]. Fourth, experienced data-science professionals including data miners, data analysts, and data scientists prefer using OSS; OSS allows users to edit code for their needs [17]. OS user beliefs also differ; OS users are more likely to value the freedom associated with modifying their software [18]. Those differences, especially the ability to edit source code, are significant in an age where individuals and institutions require customizable data tools for their complex problems. Experienced data scientists will require the use of varied tools, as there is no single machine scheme suitable to all data science deployments. As a result, OS tool development for data science occurs in numerous categories and languages. The increasing number of OS data science tools reflects this trend, totaling 70 since 2014 [13].

Data mining libraries (DMLIB) accounted for most of OS tools available, followed by Data mining environments (DME), Integration toolkits (INT), and BIS (Statistical analysis functionalities in business applications). The four most common OSS tool development languages are Java (34%), followed by Others (15%), C++ (11%), and Python (3%). Most OS development occurred on the GNU-GPL license (23%), Other Licenses (22%), and Apache (12%). In scientific and academic publication, the five most commonly used OS tools are: R (43%), followed by Waka (15%), Kepler (7%), LibSVM (7%), and Hadoop (4%). According to a survey conducted by KDNuggets, the 5 most popular programming languages (among the data science community) in 2017 were: Python (52.6%), R (52.1%), SQL (34.9%), RapidMiner (34.9%), and Excel (28.1%, although Excel is not open source) [17].

Users with less technical tasks (such as creating visualizations) prefer the use of closed-source software like Excel. OS languages like R and Python are especially used in the private sector for data-related projects. For example, research scientists at Google use R to understand "trends in ad pricing and for illuminating patterns in the search data it collects" [22]. Spotify and Netflix use open source Python modules such as Luigi for data analysis. More than half of all publicly traded companies use some form of open-source software. Increasingly, businesses no longer want to simply use OS data-science tools but be involved in development [19].

Examples of this shift has occurred at companies like Google. Historically, Google tended to be protective of their software and only released publications describing their technologies, rather than the source code [20]. Google Lab's papers, such as "MapReduce", lead to the creation of innovative OS projects like Hadoop [20]. Hadoop's popularity has possibly served as an omen to Google. In 2015, Google released TensorFlow, an open-source software library for dataflow programming (often used in machine learning and neural networks research).

Already, it is among the top 10 tools used by the data science community, experiencing a 195% increase in usage in 2017 [21]. Despite recent OS advancements, proprietary software still holds relevance within the data science field.

OSS has become ubiquitous in the security industry, but amongst data scientists, proprietary software is still used to a large extent [22]. Although OSS is dominant among professional data scientists and researchers, proprietary software is widely used in large organizations (be it governments or private companies) for various reasons. To illustrate this, it is useful to compare some features of proprietary software (SAS) and OSS (R) - both languages are used heavily in statistics [22]. While R is very popular in academia, SAS (Statistical Analysis System) has typically been dominant in the private sector. SAS can do many of the same functions as R, including: altering and retrieving statistical data, applying statistical functions and packages, and displaying impressive graphics. For small and medium sized datasets (mostly used by startups, researchers, and individuals), R performs well as a "*data analysis performer and graphics creator*" [22, 23].

For massive datasets, however, SAS offers several distinct advantages over R. R was designed for statistical computing and graphics, so "*data management tends to be time consuming and not as clean as SAS*". Students who have solely used R have an unrealistic expectation of the state of the data they receive. Furthermore, for realistic datasets (which are often messy and rarely clean), performing data manipulation in SAS is common while in "R [it] is not standard" [24]. The quality and levels of documentation also differ. R offers more open documentation than SAS, but documentation for "[R] is not well organized and often has not been thoroughly tested" [23]. Technical documentation for SAS is also vastly more detailed than R [23]. A lack of proper documentation plagues most of OSS, not just R. According to a 2017

OS survey by Github, "93% of respondents noticed that incomplete or outdated documentation is a pervasive problem" [24].

Technical support is another key differentiator, particularly among non-technical users and corporations. OSS like R relies on free support from the community; as a result, users cannot expect detailed and accurate answers in a timely manner. In contrast, SAS offers technical support to address user needs quickly [23]. Although SAS's initial costs may deter some users (the Analytics Pro version of SAS is priced at $8,700 for the first year), most of SAS's products and services offer do not charge extra for technical support. Performance differences between OS and proprietary software exists as well. Proprietary hardware drivers are typically built in "close cooperation of the hardware vendor and thus, they perform better" (Optimus Information Inc., 2015). However, optimization has its disadvantages. A common criticism of SAS is that "proper SAS does not run on Macs" [23]. While the Macintosh and Windows interfaces look different, R offers "very similar functionality" [25]. Developers are increasingly adopting Mac and Linux operating systems, resulting in more library/package support in R (as published by Stack Overflow, 2016). OS data science tools are interchangeably used in other fields, particularly AI. As the field of AI gains prominence, it is important to discuss the merit of recent advancements. What is the trend of OS tool development for AI? The next section attempts to answer this question.

**1.4 Open-Source and AI**

In 1956, RAND Corporation and Carnegie Mellon University developed the first AI language, Information Processing Language (IPL). IPL pioneered several new concepts at the time, including dynamic memory allocation, data types, recursion, and lists. However, IPL was soon replaced by a language with far simpler and scalable syntax; Lisp. According to McCarthy, the creator of Lisp, Lisp was designed to facilitate experiments for a proposed machine to "draw immediate conclusions from a list of premises" and "exhibit common sense in carrying out its instructions" [26]. At the Massachusetts institute of Technology (MIT) research lab, interest in creating Lisp machines resulted in the creation of two companies - Lisp Machines Inc (LMI) and Symbolics. Early on, differences between LMI, founded by MIT's traditional "hackers" and the more commercially oriented Symbolics emerged. A former employee of LMI, Richard Stallman, described the conflict between the two companies as a "war" [27].

Ultimately, both companies would go on to fail "with the onset of AI winter", resulting in a sharp decline in demand for specialized Lisp machines [28]. What remained of the schism was Stallman's desire to create a free operating system, paving the way for the *Free/libre open source software (FLOSS)* movement. The launch of FLOSS would ensure continued academic interest during the AI winter - a period of "*reduced commercial and scientific activities in AI*" [28]. The Open-Source movement would lead to the creation of languages like C, while not focused on AI, "fueled its development" [29]. The Deep Blue computer, an example of symbolic AI, would be developed in C. However, the use of low-level languages like C would be limited. According to Neumann [30], principal researcher at the German Research Center for Artificial Intelligence, the "fuzzy nature of many AI problems" caused programmers to develop higher-level languages that would free programmers "from the constraints of too many technical constructions". Although Lisp had solved some of these technical issues, the semantics of Lisp would prove too difficult for Lisp to become widely popular.

Regardless, concepts pioneered by Lisp, including garbage collection and object-oriented programming, would serve the development of future higher-level languages [31]. These languages include Python and Java (Java is not fully open-source but is available for free to the public). Python and Java would emerge in the 1990s and eventually become essential for machine learning. R, which "*began as an experiment in trying the methods of Lisp implementers to build a small testbed*", eventually became a popular tool for data mining (a subfield of AI) and statistics [32]. As the AI winter ended and "the problems of hardware seemed to be under control", OS software simultaneously received increased media coverage and recognition [31]. Languages like Python [33], R, Lisp, and Java were used extensively to build the first post AI-winter tools. The first "officially" OS tool specific to AI was called Torch, it was released in 2002. Torch, a machine learning library, is based on the Lua programming language [34].

**1.5 Revolutionizing Business – Avoiding Data Silos through Open Data**

Data are required for companies to operate. Its importance is ever growing. It is the fuel that machine learning runs on. Without data, nothing is known. But what is not as clear cut and universal is how the data are treated and shared; how it is improved upon, interpreted, and what can be done with it. Companies, just like governments, are readjusting to the latest data technologies and implementing the best data strategies as they are uncovered.

Data Democratization achieved through dataset publication without copyright, patent, or restrictions allow governments to be more audit-able and accountable. They accomplish that by distributing knowledge for decision making to the public. This is not fundamentally unique to governments; a similar process of choosing and concluding can be distributed amongst many instead of being handed via a centralized hierarchy. Although in the industry instance, the public would be considered the general workforce. Publicly traded companies that must submit public financial statements for regulations is a separate case; although in some sense, that is another example where the system as a whole can function better by increasing data availability and group decision making. In this industry scenario; distributing data changes the cadence of information flow and often enables a delegation of problem solving. It prevents data disagreements through increased transparency, enables increased data agility by removing barriers, and reduces data hoarding which in its own form is a source of power-siloing! It brings about an accountability that in essence can evolve things from being *words of a medicine man* to *validated knowledge of physicians,* ultimately enabling the wisdom of the many to be utilized throughout the organization.

A Data Silo can be defined as a repository of information that is controlled and accessed by a single group or department. They come in many flavors; from files sitting on computers to large databases tightly held. The most defining feature is how access is granted and how information ultimately flows. It is important to understand the different causes and consequences of siloing data; and the changes that occur as data move out into the open.

Data Silos will naturally arise in an organization and one group or department will often be the primary custodian and steward. This is organic, as centralizing particular data entry can improve these processes' accuracy and efficiency. Most companies will have a single or small number of departments that are ultimately responsible for handling accounting records and human resource records. Many will choose to have internal and external review processes where light is brought onto the accounting books. A well known example and pioneer is Google, which started publicly providing annual diversity reports in 2014 and recently released an update this year, 2019 (The Equal Pay Act - No. 899 of 2008). Through this program the company has both become aware of some of its own biases and has helped improve its hiring processes to ensure that talent is not being overlooked. It's not to say the program's goal was having the body of employees match that of the geographical area; but to ensure that hiring practices would not be influenced by unconscious factors that ultimately removed good candidates. They cross analyzed performance vs. demographics and have shown a trend towards population moving towards the geographic trend without any decrease in performance. The publication has helped spur on other companies to be able to follow suit by both raising awareness and by creating metrics that can be tracked, although many keep the data strictly for internal use.

When data is isolated and inaccessible but required; redundant data will often result. In an e-commerce company, an accounting group, a website product team, and a marketing group all require knowledge of sales performance to perform the job. Data need to be shared across the organization for such major tasks. Data Silos can also arise for a lack of technical aptitude or experience. Centralized file stores and cloud file shares have made it easier for companies to distribute documents. Spreadsheets are still the dominant data entry and processing form. They also still leave challenges around versions and editing; many workers at any business will experience at least a few emails asking for the latest version of an excel file. Stale files can result to similar disagreements as that of the redundant processing: Risks of incorrect conclusions; distrust and wasted time for validation, inefficient sharing, and ineffective processing.

Oftentimes software will use its own data file silo, such as an excel spreadsheet. Raw data is naturally sheathed by a transport and record format. Often it resorts from use of web requests hitting APIs or queries running against data warehouses. These forms are not yet widely performable by the unskilled workers; however, many tools exist to break down barriers and instill an ease of use. Another technical limitation can be the sheer size of datasets. While computing resources available have improved, network speeds have improved too, the storage medium has grown faster. It is cheap enough that many companies have chosen to capture information and left handling it later as an unsolved challenge.

Opening data to the customer base cases strong market trends and product preferences. Sharing of information to consumers has been proven vital in multiple areas and is being enforced through legislation. Food safety standards for instance, often include provisions that set up reports and databases of inspection violations. The effectiveness of transparency improving safety is also found with the publicized car safety reviews and crash statistics (although a car dealership trying to see a car with crash history would not benefit from that). The shared data aligns the public interests with market performance. The issuing of reviews and ratings has also helped infuse product confidence or desirability among consumers. It is one of the examples of online shopping that is an enhancement over the parallel in-store experience. The increase in data availability has greatly changed the knowledge dynamics between a company and its customers; and can successfully be used to improve performance; whether it be sales, safety, or satisfaction.

Companies also open up datasets to the general public to share knowledge and foster innovation. A recent surge in machine learning has caused an increase in data demand for training. By cooperating; companies have been able to improve modeling testing by having common datasets. Google recently updated its openness message:

We hope that the exceptionally large and diverse training set will inspire research into more advanced instance segmentation models. The extremely accurate ground-truth masks we provide rewards subtle improvements in the output segmentations, and thus will encourage the development of higher-quality models that deliver precise boundaries. Finally, having a single dataset with unified annotations for image classification, object detection, visual relationship detection, and instance segmentation will enable researchers to study these tasks jointly and stimulate progress towards genuine scene understanding (Google 2019).

The opening of data can also distribute decision making. I've been part of a company that automated the product experimentation tests. Prior to having shared automated numbers, longer non-repeatable analysis were performed. That required more effort from the engineering teams, causing enough friction to make testing prohibitively expensive and fairly infrequent. Automating the experimentation allowed the company to perform multiple concurrent tests with much shorter testing windows. This increased testing bandwidth helped transition data driven product management; letting everyone focus on hypothesis testing conditions while providing transparent conclusions. Decisions that were prior gut-feels on core e-commerce conversion flows could now be validated. It allowed the company to become aware of not just single sampled statistics but trials conducted over time to reassert business assumptions. It allowed product ideas to flow from the entire team; removing the organization hierarchies from leading decisions to business impact being the chief driver.

Opening data comes with challenges. There is also a balance between security and privacy that prohibits full openness with some data sources. Personal information such as health records and tax identification often remain in secured silos; and aggregations or redaction occurs to help anonymize data by removing sensitive information. Some data would jeopardize business function; it could risk employees privacy, provide enough information to allow competitors to gain an advantage, pose a legal risk or violation, and so on. Open data does not also guarantee accuracy. With the growing importance of product reviews; companies have been found to partake in false reviews on merchandise, both positive in their favor and negative on competitors' brands. Despite this, we see the overall trend that information is cheaper to be stored and captured over time; large value can be gained by opening it up.

**1.6 Future Prospects of Open Data and Open Source in the US**

Compared to the Industrial Revolution, current technological advancements are estimated to be ten times faster [35, 36]. With the advent of the information age, humans can collaborate with greater ease than before. Thomas L. Friedman, author of *The World Is Flat,* describes our new "flat world"—an ecosphere of innovation and wealth fostered by globalization [37]. A closer look at recent trends, however, does not reflect an era of lightning fast technological advancements. Federal research and development (R&D) as a percent of GDP has declined from an average of 1.2 percent in the 1970s to 0.7 percent in the 2000s [38]. In 2016, China received more patent applications than the US, Japan, South Korea, and the EU combined [35, 36]. The number of patent applications does not directly correlate with an upsurge in innovation. More balanced metrics (e.g., the number of patents per researcher), however, support the notion of declining scientific output in the US [40]. A lack of technological progress has reflected on economic performance. Economic studies indicate that as much as "85% of measured growth in US income per capita was due to technological change" [40, 41]. Yet, the average income for 'millennials' (the cohort born between 1980 and 2000) has declined from the previous generation [42]. Private investments into innovative sectors of the economy have also been lackluster. Before the financial crash of 2007, trillions of dollars flowed into a robust US economy—most of it went into government bonds and housing. Developed nations like the US depend on exporting technologically advanced products. Yet in categories of advanced technology products, including aerospace, biotechnology, and information technology, the US "has turned from a net exporter to a net importer" [35]. The percentage of Americans graduating college, particularly in science fields, is not meeting the demands of an information age. The United States has one of the lowest shares of degrees awarded in science fields [42]; 54% of all patents are awarded to foreign born students or researchers who face increasingly stringent visa restrictions [43].

Technological advancements have been dictated by the availability of data. Data drives knowledge, and data is the subject of many upcoming conflicts around the world. As Batarseh [44] mentioned in an article: "Knowledge however, has a peaceable face; like music, love and beauty – knowledge is the most amiable thing. It transfers through borders, through time, and even some preliminary studies show that knowledge can transfer via our DNA through generations. In many cases, knowledge would either hinder or thrust human progress. Fortunately, our collective knowledge is always growing. One cannot undermine the *Knowledge Doubling Curve*, it dictates the following: Until year 1900, human knowledge approximately doubled every century; by 1950 however, human knowledge doubled every 25 years; by 2000, human knowledge would double every year. Today, our knowledge is almost doubling every day! Although hard to measure or validate; as a result of such fast pace, three significant questions are yet to be answered: how are we going to manage this mammoth knowledge overflow? What if that knowledge can be organized, structured, and arranged in a way that can allow its usage? What if it falls in the wrong hands?" Well, the question remains, what if data falls hostage into *ANY* hands? And who has the authority to own data? This book, its authors, and the future of our planet dictate that no one should, a data democracy through open data and open source software is the only peaceable path forward.

43) Phung, A., 'Made in America: How Immigrants Are Driving U.S. Innovation', A Report in ThinkBig, 2012.

44) Batarseh, F., 'Thoughts on the future of human knowledge and machine intelligence', An Article on the London School of Economics (LSE), 2018.
Page | 21